# AlCoCuFeNi HIGH–ENTROPY ALLOY NANOPARTICLE MELTING AND SOLIDIFICATION: A CLASSICAL MOLECULAR DYNAMICS SIMULATION STUDY

O.I. Kushnerov

*Oles Honchar Dnipro National University, Dnipro, Ukraine*
*e-mail: kushnrv@gmail.com*

**The processes of melting and solidification of AlCoCuFeNi nanoparticle of about 10 nm is studied by molecular dynamics simulation at three different cooling rates ($1 \cdot 10^{11}$ K/s, $1 \cdot 10^{12}$ K/s, and $1 \cdot 10^{13}$ K/s) using the embedded atom model (EAM) potential. The melting and crystallization of the nanoparticle are characterized by studying the temperature dependence of the potential energy. The adaptive common neighbor analysis (CNA) is performed and the radial distribution function (RDF) is calculated to determine the structure and lattice parameters of phases of the AlCoCuFeNi nanoparticle. It is shown that the final structure of the investigated nanoparticle changes from amorphous to crystalline with decreasing of the rate of cooling, and the temperature hysteresis takes place during the melting and crystallization of AlCoCuFeNi HEA nanoparticle.**

## 1. Introduction

The high-entropy alloys (HEAs) or multi-principal component alloys have been widely studied since 2004 [1]. Such alloys usually contain from 5 to 13 major elements in equiatomic or near equiatomic concentrations (5 to 35%). Because of the high entropy, usually simple substitutional solid solutions with BCC or FCC crystal lattices are formed during the solidification of multicomponent alloys. The HEAs are characterized by unique structures and a number of useful characteristics, such as hardness, wear-resistance, resistance to oxidation, corrosion, and ionizing radiation, high thermal stability and biocompatibility [1–3]. Thus, the HEAs show promise as materials for application in electronics, atomic power engineering, transportation equipment, space-rocket hardware, medicine, etc. Nanoparticles are defined as small clusters of material, typically less than 100 nm in diameter. Nanoparticles offer many potentially interesting mechanical and chemical properties that differ from the bulk material [4]. These properties arise because of the high surface area to volume ratio inherent at the nanometer scale. Nanoparticles can also act as a bridge between the continuum and atomistic length scales because they contain a relatively small number of atoms. Some experimental observations of nanoparticles are very difficult because of the short length and time scales involved. In order to quantify the properties of individual nanoparticles computer simulation can be used in conjunction with experimentation to obtain reliable data and conclusions. Atomic interactions in a metallic virtual material are realistically represented with the use of quantum mechanics through a time-dependent Schrödinger equation. At the same time, it has been shown that molecular dynamics simulations based on classical mechanics were able to reproduce with sufficient accuracy the experimental results of deformation mechanisms and atomic motions in metals and to provide an atomic description of the crystallization and glass-forming processes during the rapid solidification of alloys. Also, molecular dynamics is one of the most powerful atomic simulation methods that are used to investigate the properties and structures of nanomaterials.The present study is concerned with classical molecular dynamics (MD) simulation for studying the melting and solidification of AlCoCuFeNi HEA nanoparticle.

## 2. Simulation model and methods

### 2.1. Potential model

For describing the interaction between Al–Co–Cu–Ni–Fe the EAM potential from [5] is used. The embedded-atom method (EAM) was created by Daw and Baskes and describes the interaction between atoms in metals and intermetallics [6]. EAM potentials involve empirical formulas that simulate material characteristics such as the heat of solution, lattice constants, and surface energies. Such a potential is non-pairwise in the sense that it is based on concepts from density functional theory, which stipulate in general that the energy of a solid is a unique function of the electron density.

In the embedded atom method (EAM) the energy consists of two terms: one term describes the delocalized electrons by a many-body term as an embedding energy function $f_i(\bar{\rho}_i)$ of the local host electron density $\bar{\rho}_i$, the second term is a pairwise additive term $\phi_{ij}(r_{ij})$ of the interaction between atomic cores [6]. Normally, the published EAM potentials for atoms of a single element are based upon different reference states and the range of interaction cut-off distances. Hence, the functions $f_i(\bar{\rho}_i)$, $\bar{\rho}_i$, and $\phi_{ij}(r_{ij})$ fitted for elemental metals cannot be directly applied to alloy systems. But the alloy EAM potentials have been made from elemental EAM potentials by normalising them and introducing an EAM alloy model. Such alloy potentials have been employed in MD simulations to generate results consistent with experimental observation [5]. In the EAM alloy model, the total energy of a binary system A–B is represented as

$$E_{pot} = \sum_i f_i(\bar{\rho}_i) + \frac{1}{2}\sum_i \sum_{j\neq i} \phi_{ij}(r_{ij}) \tag{1}$$

Here, $\phi_{ij}(r_{ij})$ is a pair interaction potential energy as a function of the distance $r_{ij}$ between atoms $i$ and $j$ that have chemical sorts $a$ and $b$, and $\bar{\rho}_i$ is given by

$$\bar{\rho}_i = \sum_{j\neq i} \rho_j(r_{ij}) \tag{2}$$

where $\rho_j(r_{ij})$ is the electron density function assigned to an $j$ atom. The functional form of pair potentials are written as

$$\phi(r) = \frac{A\exp\left\{-\alpha(\frac{r}{r_e}-1)\right\}}{1+(\frac{r}{r_e}-k)^{20}} - \frac{B\exp\left\{-\beta(\frac{r}{r_e}-1)\right\}}{1+(\frac{r}{r_e}-\lambda)^{20}} \tag{3}$$

where $A$, $B$, $\alpha$ and $\beta$ are four adjustable parameters, $r_e$ is an equilibrium distance between nearest neighbors, $k$ and $\lambda$ are cut-off parameters. The electron density function is similar to the second term in the pair potential but $f_e$ is used instead of $B$, i.e.

$$\rho(r) = \frac{f_e\exp\left\{-\beta(\frac{r}{r_e}-1)\right\}}{1+(\frac{r}{r_e}-\lambda)^{20}} \tag{4}$$

The pair potential between different species $a$ and $b$ is then constructed as

$$\phi_{ab}(r) = \frac{1}{2}\left[\frac{f_b(r)}{f_a(r)}\phi_{aa}(r) + \frac{f_a(r)}{f_b(r)}\phi_{bb}(r)\right] \tag{5}$$

The embedding energy functions require that three equations are used to fit three different electron density ranges. These equations are

$$f(\bar{\rho}) = \sum_{i=0}^{3} F_{ni}(\frac{\rho}{0.85\rho_e} - 1)^i \text{ , } \rho < 0.85\rho_e \tag{6}$$

$$f(\bar{\rho}) = \sum_{i=0}^{3} F_i(\frac{\rho}{\rho_e} - 1)^i \text{ , } 0.85\rho_e \le \rho < 1.15\rho_e \tag{7}$$

$$f(\bar{\rho}) = \sum_{i=0}^{3} F_n \left[1 - \eta \ln(\frac{\rho}{\rho_s}\right](\frac{\rho}{\rho_s})^{\eta} \text{ , } \rho \ge 1.15\rho_e \tag{8}$$

where $F_{ni}$, $F_i$, and $F_n$ are tabulated constants, $\rho_e$ is an equilibrium electron density [5].

### 2.2. Computational details

Classical MD simulation studies of high entropy AlCoCuFeNi nanoparticle have been performed using Large Scale Atomic/Molecular Massively Parallel Simulator (LAMMPS, Sandia National Laboratory, USA) [7]. Visualization of snapshots and the coordination analysis are done using open visualization tool software OVITO [8]. Simulations are performed under the NVT ensemble (constant number of atoms, the volume of the main cell and temperature), but the sample volume was not constant due to the fact that the sample was with a free surface (nanoparticle was smaller than the main cell). Integration was done by the velocity-Verlet algorithm with a timestep of 0.002 ps. The Nosé–Hoover thermostat was used in the MD simulation to perform temperature control. The simulated nanoparticle was composed of 50000 atoms (Al, Co, Cu, Fe, Ni in the equiatomic ratio) and the average size of nanoparticle was ~ 10 nm. The system was heated up to 2300 K and subsequently annealed at this temperature for 200 ps. After this, the nanoparticle was quenched from a molted state at a rate of $1 \cdot 10^{11}$ K/s (low), $1 \cdot 10^{12}$ K/s (medium), and $1 \cdot 10^{13}$ K/s (high) to 300 K. After quenching the radial distribution function (RDF) was calculated for determining the structure of the nanoparticle. Also, the adaptive common neighbor analysis (CNA) was performed to identify the microstructure of the samples (on the basis of an algorithm determining the symmetry of the nearest-neighbor atom). The AlCoCuFeNi nanoparticle cooled to a temperature of 300 K at a low rate was then heated up to 2000 K with the rate of $1 \cdot 10^{11}$ K/s to study the melting process.

## 3. Results and discussion

The potential energy variations of AlCoCuFeNi nanoparticle at different cooling rates are shown in Fig. 1. The exact melting point could be calculated by simultaneously equilibrating the solid and liquid phases in a simulation box on the macroscale. But in the nanoscale, the transition from solid to liquid phase could be identified by a jump in the energy curve (Fig. 1) [9]. Apparently, different cooling rates lead to different features. At cooling rates $1 \cdot 10^{12}$ K/s and $1 \cdot 10^{13}$ K/s, the potential energy decreases almost linearly with the temperature decreasing. However, there is a drop in potential energy at the cooling rate of $1 \cdot 10^{11}$ K/s. The reasons for this phenomenon may be explained by atomic diffusion. At high cooling rates ($1 \cdot 10^{12}$ K/s, $1 \cdot 10^{13}$ K/s), the atoms have not enough time to rearrange their position in the cooling process, therefore, the potential energy curve varies almost linearly with the temperature, and the final structure is amorphous. At low cooling rate ($1 \cdot 10^{11}$ K/s), the atoms have enough time to rearrange the positions, and then the crystallization happens in the range of 885–855 K. It should be noted that at medium rate ($1 \cdot 10^{12}$ K/s), the atoms have some time to rearrange their position to obtain the more compact structure with lower values of potential energy (Fig. 1a). At the Fig. 1b the potential energy dependencies are given for the solidified and subsequently melted nanoparticle. Variations of the curve slopes in this diagram show liquid-solid (from 885 to 855 K) and solid-liquid (from 1220 to 1265 K) phase transitions. The differences in the melting and crystallization temperatures can be explained by the temperature hysteresis that takes place during the melting and crystallization of nanoobjects [10, 11]. When approaching the melting temperature, the thin surface layer of the nanoobject first melts, then with a further increase in temperature the thickness of the liquid surface layer increases [11]. Thus, the new phase nucleates at a temperature below the melting point, and this circumstance prevents the existence of a nanoobject in the solid phase above the melting temperature, i.e. prevents overheating of the nanoobject. The observed hysteresis between crystallization and melting of nanoobjects is due to the fact that the size of the critical nucleus of a new phase exceeds the characteristic size of the nanoobject itself [10, 11]. This makes it impossible to form solid phase nuclei in a certain temperature range below the melting point. Thus, the crystallization temperature of nanoobjects can be significantly lower than their melting temperature. For the present MD simulation studies of high entropy AlCoCuFeNi nanoparticle the ratio between the temperatures is $T_{cryst} \approx 0.7T_{melt}$, which is typical for nanoobjects with a size of 10 nm or less.

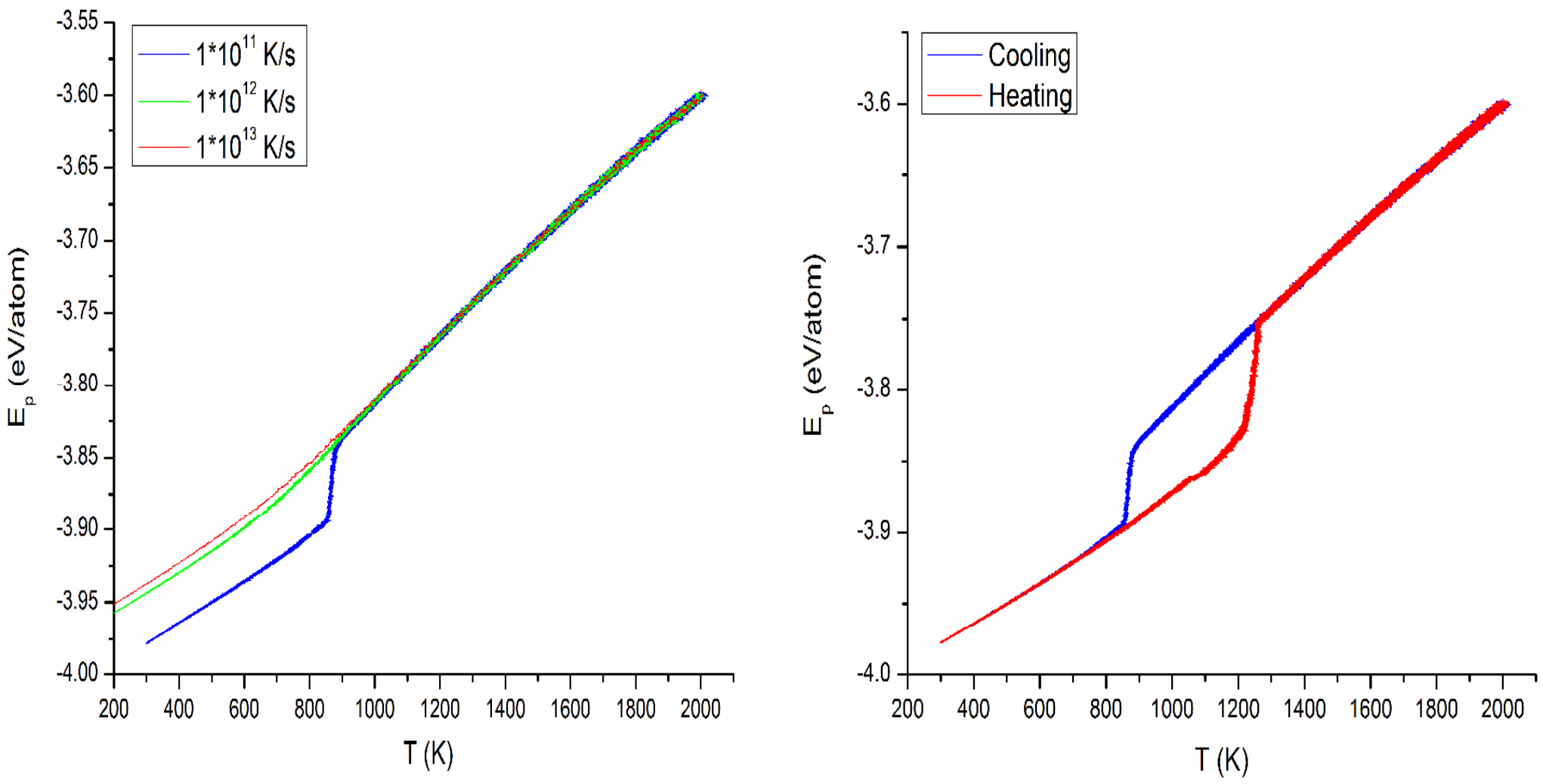

**a** **b**

**Fig. 1. The temperature dependencies of the potential energy per atom of AlCoCuFeNi nanoparticle.**

The RDF curves (at the temperature of 300 K) for AlCoCuFeNi nanoparticle crystallized at a rate of $1 \cdot 10^{11}$ K/s are given in Fig. 2a. By the CNA analysis, it has been established that the simulated nanoparticle contains the FCC phase (59.4 %), BCC phase (11.1 %), HCP phase (10.5 %) and the unrecognized (Other) phase (19%), which, in accordance with RDF, has an amorphous structure. The estimated from the RDF curves [12] lattice parameter of the BCC phase for the present MD simulations is 0.291 nm, the FCC one is 0.367 nm and for the HCP phase $a$=0.258 nm and $c$=0.447 nm. These values are in a good agreement with experimental data [13] where for casted AlCoCuFeNi alloy at room temperature the lattice parameters were 0.2878 nm and 0.3624 nm for the BCC and FCC phases, respectively. The features of the distribution of the HCP phase atoms (Fig. 3) allows us to conclude that this phase consists partly of stacking faults (a local deviation of the stacking sequence of atomic layers) in the lattice of the FCC phase, and partly of twin boundaries between the FCC twins [14]. As for the amorphous phase, its appearance can be explained by the high cooling rate at which some of the atoms have not enough time to rearrange and create a crystalline phase.

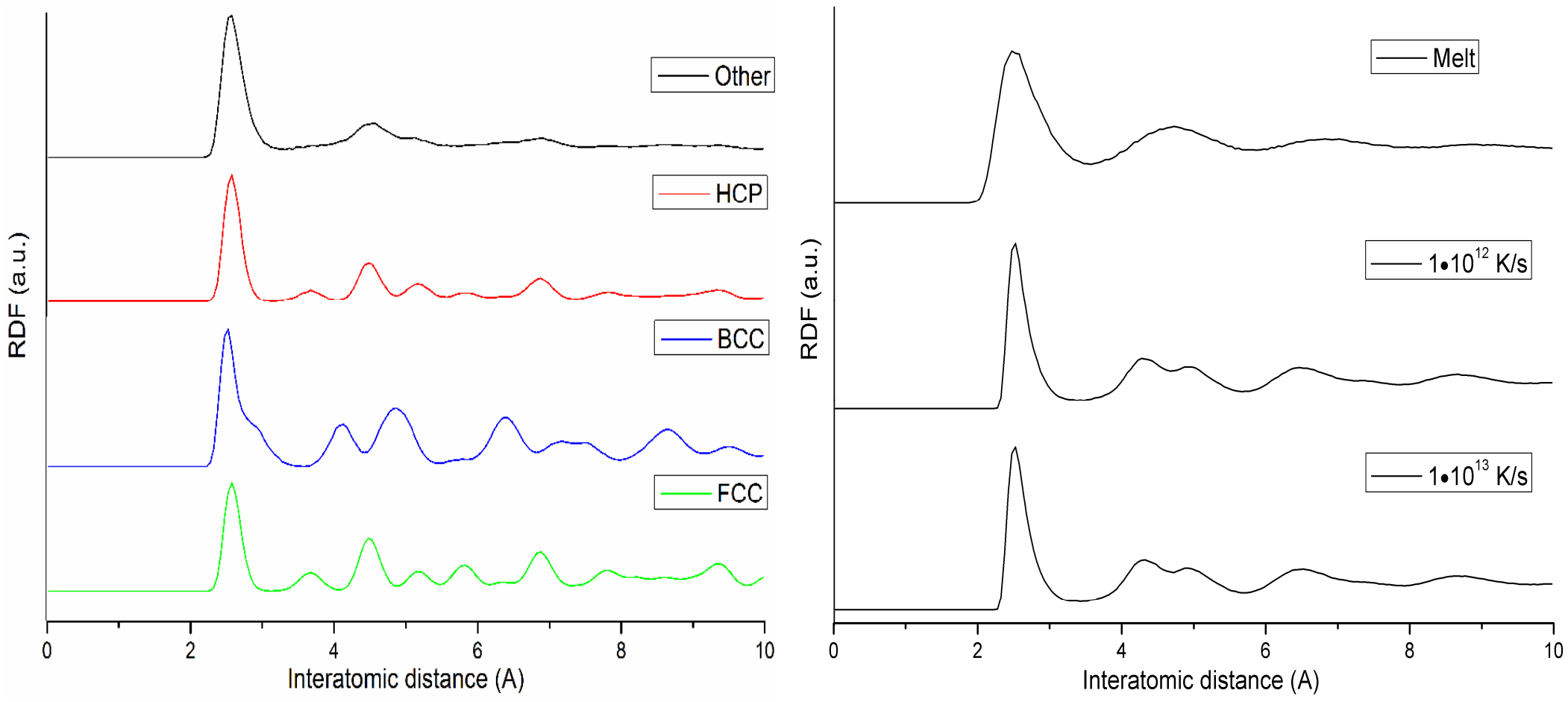

**Fig. 2. Total RDFs for AlCoCuFeNi nanoparticle: a– crystallized at a rate of $1{\cdot}10^{11}$ K/s, individual curves correspond to different phases; b – cooled to a temperature of 300 K at different rates.**

Total RDFs for cooled to a temperature of 300 K at different rates AlCoCuFeNi nanoparticles are shown in Fig. 2b. The curve for the initial molten state is also given for comparison. Both for the $1{\cdot}10^{12}$ K/s and for the $1{\cdot}10^{13}$ K/s cooling rates the second peak of the RDF curve has the tendency to split, which suggests that the metallic glass has formed (in contrast to the curve for the initial liquid) [15].

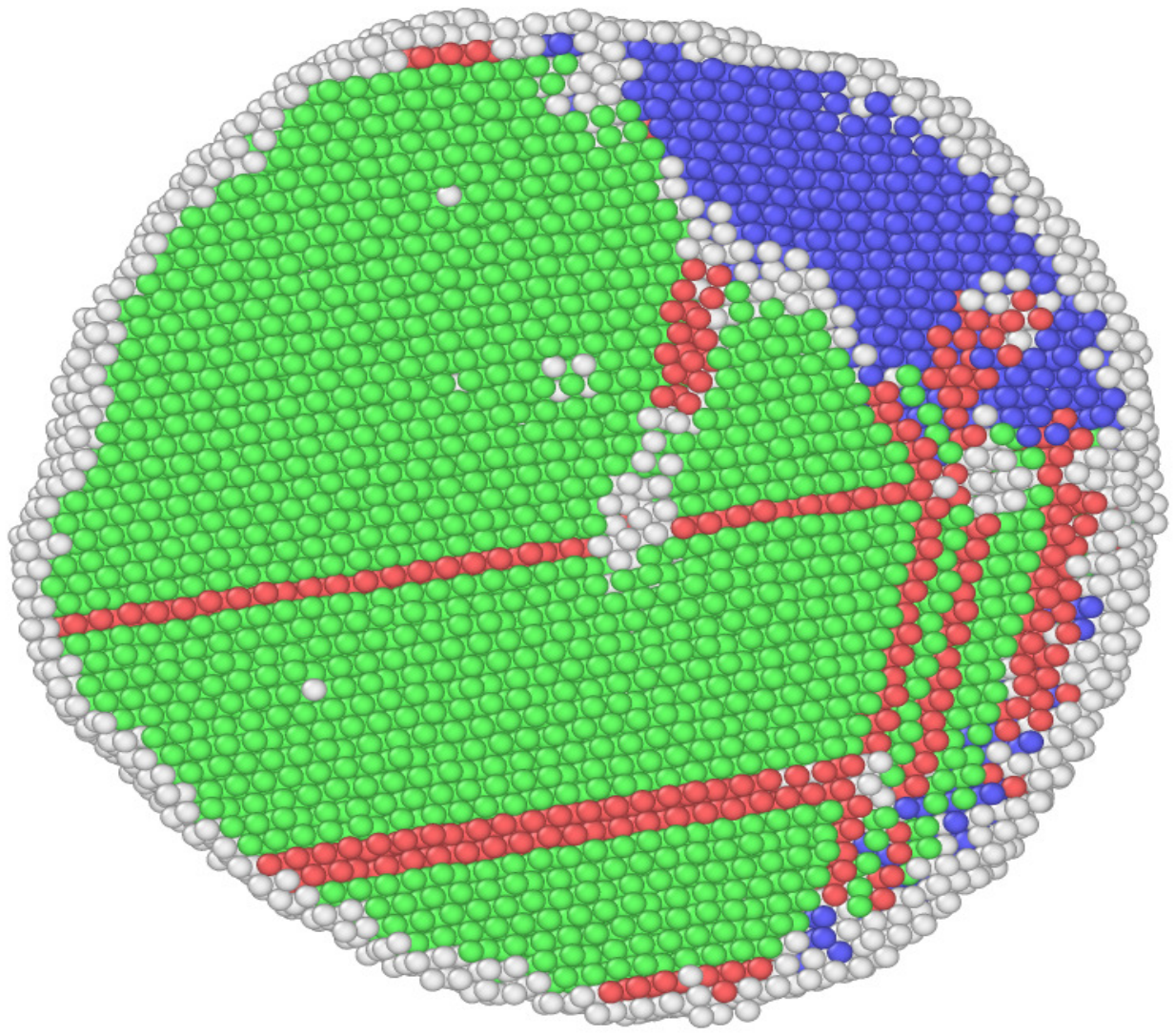

**Fig. 3. Cross-section of AlCoCuFeNi nanoparticle:**
**● – FCC, ● – BCC, ● – HCP, ● – amorphous phase.**

## 4. Conclusions

The melting and solidification processes of AlCoCuFeNi nanoparticle of about 10 nm has been studied by molecular dynamics simulation at three different cooling rates. It is found that the final structure of investigated nanoparticle changes from amorphous to crystalline with the rate of cooling processes decreasing. From the simulation results, it is shown that the classical molecular dynamics simulation with the use of embedded atom potential allows us to describe structure and lattice parameters of AlCoCuFeNi nanoparticle in good agreement with experimental data both for the as-casted alloy and for thin melt-spinning ribbons. It is shown, that the temperature hysteresis takes place between the melting and crystallization of AlCoCuFeNi HEA nanoparticle.